\documentclass[aps,pre,twocolumn, superscriptaddress]{revtex4}

\usepackage{graphicx}
\usepackage{amsmath}
\usepackage{color}

\begin{document}
\title{The Role of Quantum Fluctuations in the Hexatic Phase of Cold Polar Molecules}

\author{Wolfgang Lechner}
\email{w.lechner@uibk.ac.at}
\affiliation{Institute for Quantum Optics and Quantum Information, Austrian Academy of
Sciences, 6020 Innsbruck, Austria}
\affiliation{Institute for Theoretical Physics, University of Innsbruck, 6020 Innsbruck,
Austria}

\author{Hans-Peter B\"uchler}
\affiliation{Institut f\"ur Theoretische Physik III, University of Stuttgart, Germany}

\author{Peter Zoller}
\affiliation{Institute for Quantum Optics and Quantum Information, Austrian Academy of
Sciences, 6020 Innsbruck, Austria}
\affiliation{Institute for Theoretical Physics, University of Innsbruck, 6020 Innsbruck,
Austria}

\date{\today}
  
\begin{abstract} 
Two dimensional crystals melt via an intermediate \textit{hexatic} phase which is characterized by an anomalous scaling of spatial and orientational correlation functions and the absence of an attraction between dislocations. We propose a protocol to study the role of quantum fluctuations on the nature of this phase with a system of strongly correlated polar molecules in a parameter regime where thermal and quantum fluctuations are of the same order of magnitude. The dislocations can be located in experiment from local energy differences which induce internal stark shifts in dislocation molecules. We present a criterium to identify the hexatic phase from the statistics of the end points of topological defect strings and find a hexatic phase, which is dominated by quantum fluctuations, between crystal and superfluid phase.
\end{abstract}
\pacs{67.85.-d, 61.72.Ff} 
 
\maketitle

With the remarkable recent experimental progress in manipulating ensembles of molecules, it is now possible to reach the micro-Kelvin temperature regime with the prospect to study the dynamics of ultracold, strongly correlated quantum systems \cite{REVIEWCEHM,CARR}. Trapping techniques allow for the realization of long range dipolar systems in two dimensions \cite{KKNI,OSPELKAUS,Yan,LEV,HCN} and in particular molecules with large dipole moments (e.g. LiCs \cite{WEIDEMUELLER} or NaK \cite{NAK}) may exhibit the formation of self-assembled dipolar crystals \cite{Baranov, Lewenstein, Buechler,LEWENSTEINBOOK,Cooper, Astra,Z_PRL_SELFASSEMBLY}. Two dimensional crystals are fundamentally different compared to their three dimensional counterparts, as was first noted by Peierls \cite{PEIERLS} and Landau \cite{LANDAU}. They conjectured that in two dimensions there is no true long range order but only quasi-long-range order \cite{MERMINWAGNER}. In the semial paper, Kosterlitz, Thouless, Nelson, Halperin and Young (KTHNY) argued that the melting from such a quasi-crystalline phase to the liquid is a two step process with an intermediate phase \cite{KTHNY}. This so-called \textit{hexatic} phase exhibits short range translational - but quasi-long range bond-orientational order. On the microscopic level this can be understood from the formation and dissociation of dislocation pairs \cite{KTHNY,TAYLOR}. In the classical regime, the existence of the hexatic phase has been debated for decades and only recent large scale computer simulations \cite{SIMULATIONHEXATIC,KRAUTH2,KRAUTH} and experimental realizations of colloidal dipolar crystals with single particle resolution have shown evidence for the KTHNY scenario (review see \cite{MELTINGREVIEW}).  

\begin{figure}[ht]
\includegraphics[width=8.5cm]{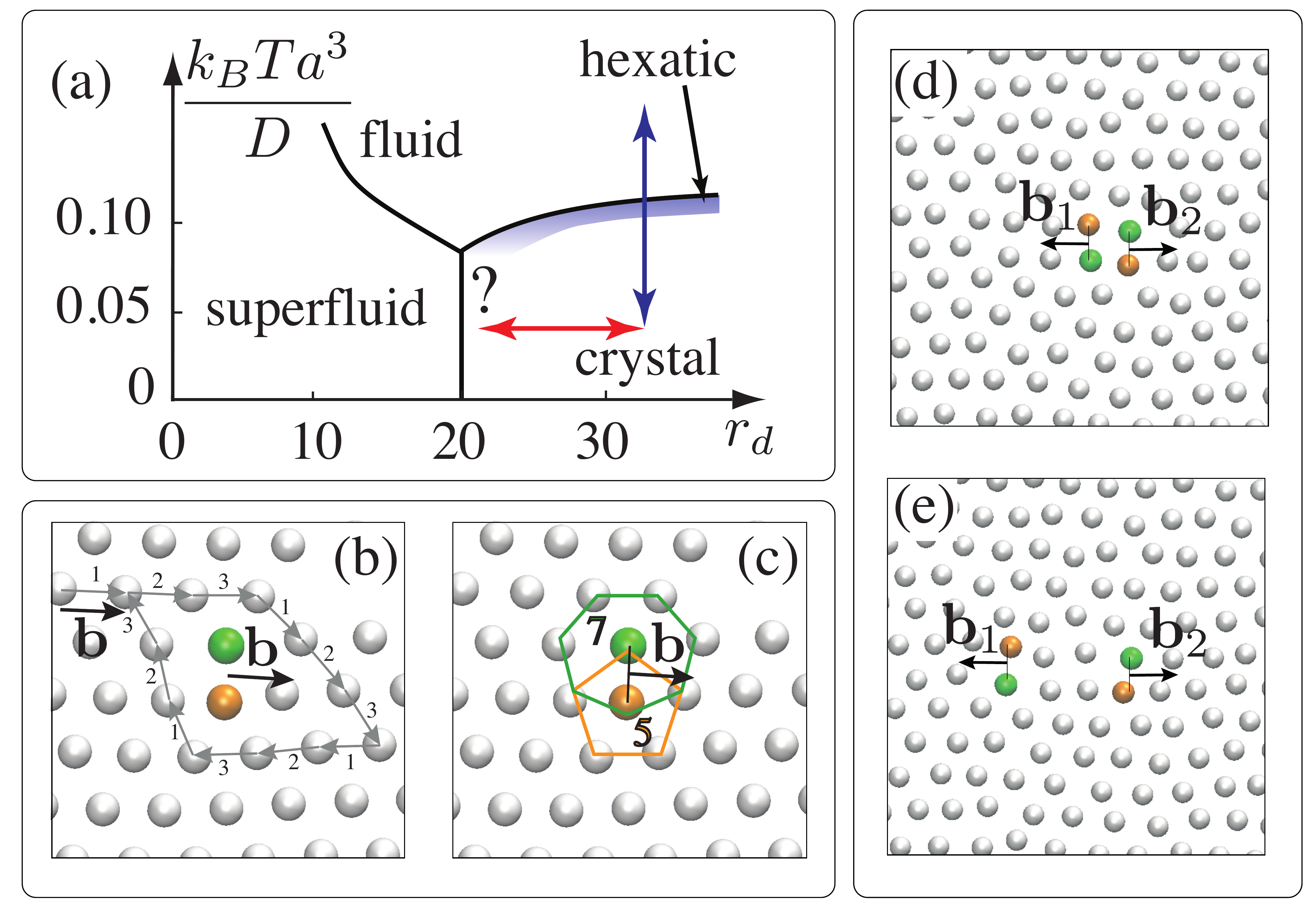}
\caption{(a) The phase diagram of a two-dimensional dipolar system exhibits a superfluid, normal liquid, crystal and an intermediate \textit{hexatic} phase. In the hexatic phase dislocation defects form in pairs and separate spontaneously. The hexatic phase can be reached from the crystal by increase of temperature (blue). Here we show that a hexatic phase can also result from quantum fluctuations (red). (b) and (c) Dislocations are characterized by a Burgers vector which can be constructed from two consistent alternative procedures: (b) By following a closed loop around a dislocation with a constant number of lattice sites in each direction the starting and end point are separated by the Burgers vector. (c) Alternatively, the Burgers vector can be constructed from a nearest neighbor analysis of the individual particles. Using a Voronoi construction \cite{VORONOI}, the number of neighbors in a perfect triangular crystal is $N_{nb}=6$ for each particle. A dislocation is defined as a pair of particles with $N_{nb}=5$ (orange) and $N_{nb}=7$ (green). The Burgers vector is orthogonal to the connection between $5$ and $7$ coordinated particles. (d) Due to thermal or quantum fluctuations, dislocations can form spontaneously in pairs with opposite Burgers vectors. (e) Separation of dislocation pairs is a rare event in the crystal phase due to strong attractive forces. The hexatic phase is characterized by the absence of this interaction leading to free pairs of dislocation. }
\label{fig:fig1}
\end{figure}

Here, we study the influence of quantum fluctuations on the nature of the hexatic phase and the dynamics and interaction of dislocations.
We propose an experimental setup with trapped cold polar molecules that allows one to tune system parameters from the classical to the quantum regime and provide theoretical evidence of a hexatic phase between the crystal and superfluid where quantum fluctuations are dominant. The relevant regime of strong correlations and temperatures where quantum fluctuations are of the same order as thermal fluctuations [see Fig. \ref{fig:fig1}(a)] can be reached e.g. with LiCs \cite{WEIDEMUELLER,LECHNERGLASS} or NaK \cite{NAK}. 
While crystal and liquid phase can be distinguished with established scattering methods (see \cite{supmat}), here, we propose a protocol that allows for the detection of the hexatic phase in our setup from the interaction between dislocations which is measured from the statistics of the end-points of defect strings \cite{LECHNERDEFECTSTRINGS}. These points are topologically protected dislocations which cannot annihilate and their interaction is of double-well shape \cite{LECHNERDEFECTSTRINGS}. We derive the critical value for the barrier height from elasticity theory which allows one to distinguish between hexatic and crystalline phase. In particular we propose a measurement protocol, employing local energy differences of dislocation particles (and therefore stark shift of internal level) between molecules in the perfect crystal and dislocations.

We assume trapped polar molecules with a strong confinement to two dimensions and a static electric field perpendicular to the plane leading to a 2D system of aligned dipoles \cite{Buechler}. For a homogeneous system, the corresponding 2D Hamiltonian is 

\begin{equation}
H = \sum_i \frac{\mathbf{p}_i^2}{2m} + \sum_{i<j} \frac{D}{r_{ij}^3}.
\label{equ:H}
\end{equation}
Here, $\mathbf{p}_i$ and $m$ are momentum and mass of the particle $i$, respectively, and $\mathbf{r}_{ij}$ is the distance between particle $i$ and $j$ in the xy-plane. The effective interaction strength is $D=d_{\text{eff}}^2/(4 \pi \epsilon_0)$, with $d_{\text{eff}}$ the induced dipole moment. The equilibrium phases of the systems are characterized by the parameters $r_d  = D m/(\hbar^2 a)$, which measures the ratio between kinetic energy and potential energy for particles with average distance $a$, and the dimensionless temperature $\kappa = k_B T a^3/D$. In the case of bosons \cite{Buechler}, the phase diagram [depicted in Fig. \ref{fig:fig1}(a)] features a superfluid, normal liquid and crystalline phase. The limit $r_d \rightarrow \infty$ corresponds to the classical limit where the hexatic phase is found at $\kappa \approx 0.1$ \cite{MARET}. Below we will show that for small $r_d$ a hexatic phase dominated by quantum fluctuations exists between crystal and superfluid phase. 

The key feature of the hexatic phase is the anomalous scaling of translational and orientational order as a result of the softening of elastic constants. Translational order is characterized by the scaling of the density-density correlation $g(r) = \langle \rho(\mathbf{r}_0+\mathbf{r})  \rho(\mathbf{r}_0) \rangle$. Here, angular brackets $\langle \cdot \rangle$ indicate the canonical ensemble average and $r=|\mathbf{r}|$. Orientational order is given by $\mbox{$g_6(r) = \frac{1}{N} \langle \sum_m \psi_m^*(\mathbf{r}_0+\mathbf{r}) \psi_m(\mathbf{r}_0) \rangle$}$, with $\psi_m(r) = \frac{1}{N_{nb}} \sum_j e^{6 \theta_{jm}(\mathbf{r})}$. Here, the sum runs over all nearest neighbors of particle $m$. In contrast to the crystalline and liquid phase, the  hexatic phase is characterized by a broken translational order $g(r) \sim \exp(-\eta r)$ and (quasi-) long-range ($g_6(r)\sim r^{-\eta}$)  orientational order. 

In the following we study how quantum fluctuations can lead to softening of the elastic constants similar to thermal fluctuations in the classical case.  In principle, the hexatic phase can be understood from the scaling of these correlations functions. However, this requires large system sizes which are not realistic in an experimental setup with cold molecules. Below, we propose a method to identify the hexatic phase in systems of cold molecules with realistic numbers of particles based on the microscopic picture of dislocation unbinding due to quantum fluctuations.

\begin{figure}[]
\centerline{\includegraphics[width=8.6cm]{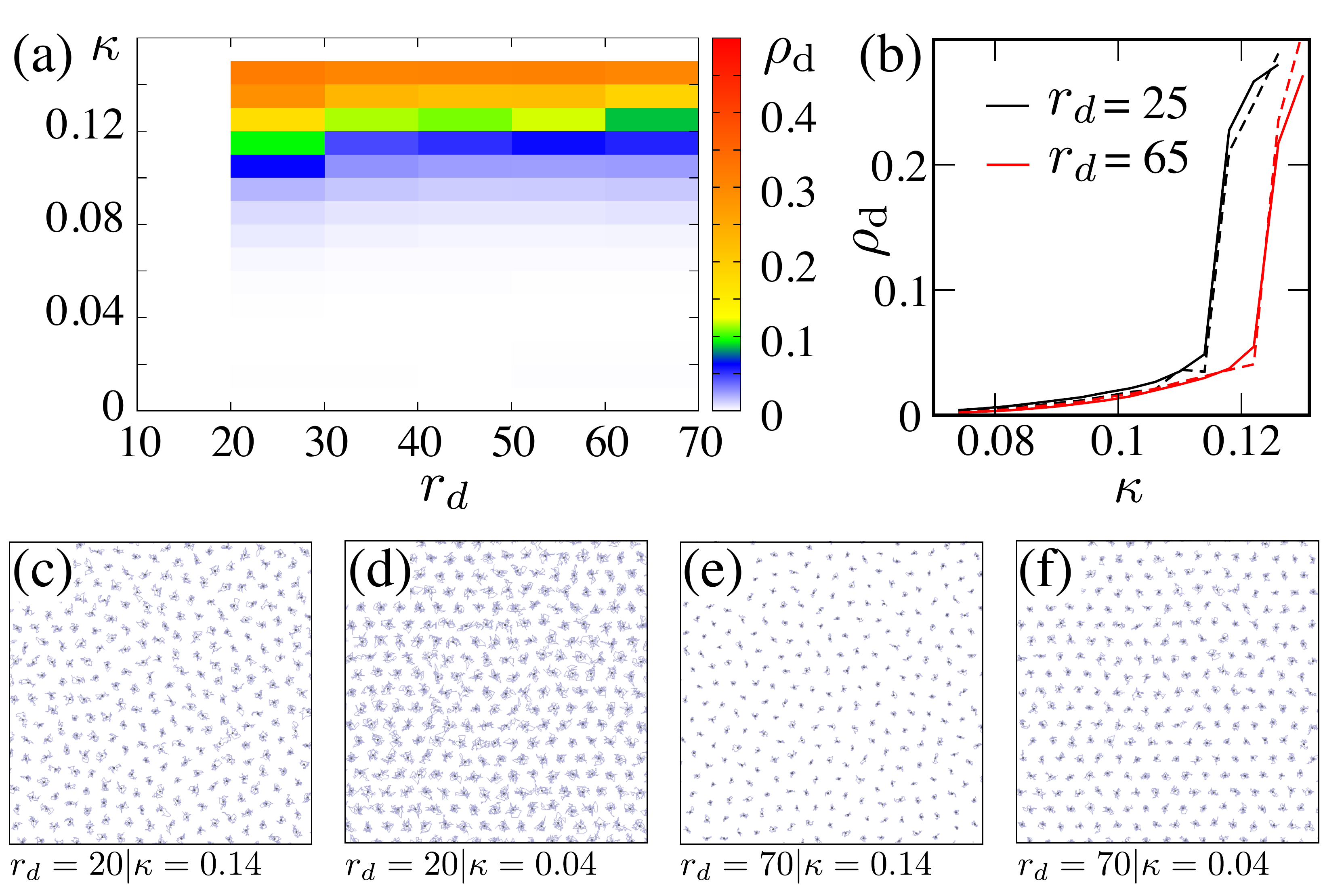}}
\caption{(a) Dislocation density in the dipolar system as a function of $r_d$ and $\kappa$. In a prefect crystal (white) the number of dislocations is vanishing while in the hexatic phase dislocations emerge (blue). Above melting the system is dominated by defects (red). By lowering $r_D$, the melting line shift due to quantum effects from $\kappa=0.12$ in the classical limit to $\kappa=0.1$. (b) Dislocation density as a function of the temperature $K$ for $r_d=25$ (black) and $r_d=65$ (red) for $N=780$ (solid) and $N=320$ (dashed). (c-f) Typical snapshots of the projected paths. (c) At high temperatures and small $r_d$, both thermal fluctuations and quantum fluctuations are large and the system melts. (d) Below the melting line, the system melts due to the large quantum fluctuations at small $r_d$. (e) At high temperature and large $r_d$ the system enters the classical liquid phase with well localized wave functions. (f) In the classical crystal phase at large $r_d$ and small $\kappa$ the system self-assembles into a triangular lattice without defects. }
\label{fig:fig2}
\end{figure}

Dislocation interactions are at the heart of the hexatic phase. The defects are specified by a position and a direction, the so-called Burgers vector. The Burgers vector can be defined in two alternative ways, as depicted in Figs. \ref{fig:fig1}(b) and (c). Dislocations can form in pairs due to thermal (and as we show below also due to quantum) fluctuations in triangular crystals. These pairs of defects, depicted in Fig. \ref{fig:fig1}(d), have opposite Burgers vectors. In such a configuration, dislocations are strongly attractive in the crystal phase and, therefore, annihilate very fast and only separate as a results of rare events [depicted in Fig. \ref{fig:fig1}(e)]. In the hexatic phase this attraction vanishes and dislocation pairs can dissociate into free dislocations. 

\begin{figure}[htb]
\centerline{\includegraphics[width=8.5cm]{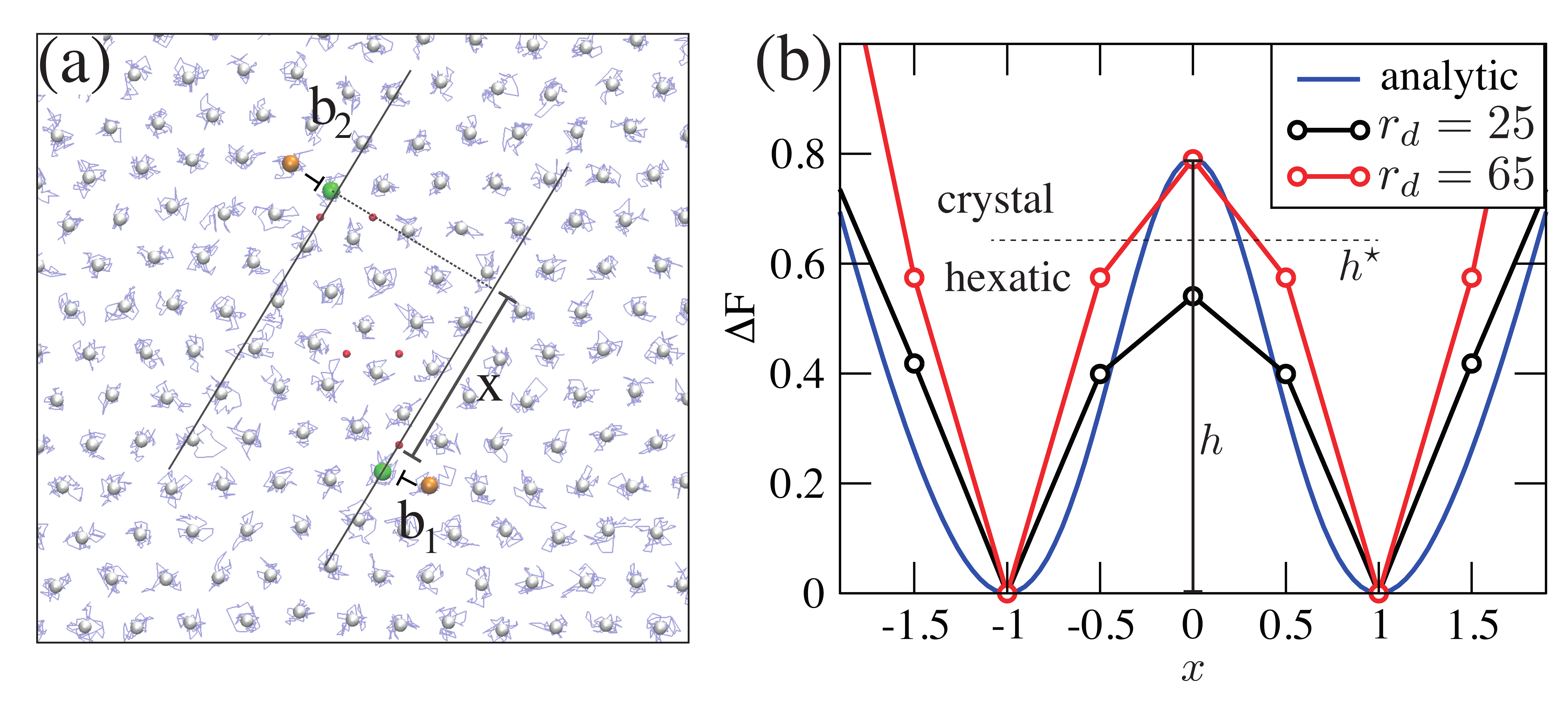}}
\caption{Measurement of the dislocation interaction from a topological defect string. (a) Typical snapshot of a string configuration with $N=5$ vacancies from a path integral simulation. The particle positions (spheres) are the averaged positions from the projected path integrals (light blue). The system consists of particles with $N_{nb}=6$ neigbhors (white), $N_{nb}=5$ neighbors (orange) and $N_{nb}=7$ neighbors (green). The dislocations of the vacancies annihilate in pairs leaving only two dislocations at the end points of the string (red dots). The motion of the dislocations is constraint by the topology of the string (details see \cite{LECHNERDEFECTSTRINGS}). (b) The separation probability in direction of motion of the string $x$ [$\kappa=0.054$ and  $r_d=65$ (red,classical) and $r_d=25$(black)] is measured from the dynamics of the path integral simulation \cite{supmat} and is compared to the analytic expression (blue) from Eq.~(\ref{equ:endtoendscaled}). From the barrier height, one can calculate the interaction strength of dislocations with Eq.~(\ref{equ:barrierheight}). Below the critical value $h<0.63$ the system enters the hexatic phase (see main text). }
\label{fig:fig3}
\end{figure}

Fig. \ref{fig:fig2}(a) depicts the dislocation density as a function of $r_d$ and $\kappa$. The time evolution of the Eq.~(\ref{equ:H}) is sampled from a semi-classical path integral simulation \cite{supmat}. This method neglects exchange and is not applicable in the deep quantum regime. Dislocations are identified from a Voronoi construction \cite{VORONOI}. The classical limit in this phase diagram corresponds to large $r_d$. In this regime, the phase boundary between liquid and solid is characterized by a single parameter $\kappa$. The classical hexatic phase is expected at $\kappa \approx 0.1$ \cite{MARET}. This is in agreement with our path integral simulations (see Fig. \ref{fig:fig2}). The crystal melts into a liquid at $\kappa=0.12$. We find a finite number of dislocations also for $\kappa<0.12$, indicating a possible hexatic phase, with a dislocation density of $\rho_d<0.05$, where $\rho_d=(N_5+N_7)/N$ is the number density of particles with $5$ and $7$ neighbors respectively [see construction of dislocations from Fig. \ref{fig:fig1}(c)]. Figs. \ref{fig:fig2}(e) and (f) depict typical snapshots of the system in this regime. The particles are well localized above and below the melting temperature and the crystal melts purely because of thermal motion.

In the regime where quantum fluctuations are relevant, the melting curve changes considerably. For $r_d<20$ the system enters a superfluid phase \cite{Buechler}. Note, that our semi-classical approach is not applicable in the superfluid region but, remarkable, does predict the crystal to superfluid transition accurately. Here, the most interesting case is the crystal region ($r_d\approx25$) close to the superfluid transition. The melting temperature decreases considerable due to quantum fluctuations as depicted in Fig. \ref{fig:fig2}(b). Also, the number of dislocations is larger compared to $r_d=65$ for all $\kappa$. Typical snapshots with large quantum fluctuations are depicted in Figs. \ref{fig:fig2}(c) and (d). The results presented in Fig. \ref{fig:fig2}(b) indicated that quantum fluctuations also have a strong influence on the dislocation formation and therefore the hexatic phase. 

In the following we present a realistic scheme to identify the quantum hexatic phase from the statistics of end-points of defect strings \cite{LECHNERDEFECTSTRINGS}. Defect strings are self-assembled strings of point defects such as vacancies or interstitials \cite{LDSoftMatter1,LDSoftMatter2}. A free point-defect consists of several dislocations. In the string, pairs of dislocations annihilate each other so that only two dislocations at each end of the strings remain. These two dislocations are protected by topology and cannot annihilate. The statistics of the end-point positions allows one to measure the interaction between dislocations in the following protocol:

According to elasticity theory, the interaction between two dislocations with Burgers vectors ${\bf b}_1$ and ${\bf b}_2$ and separation ${\bf R}$ is given by \cite{MARET} 
\begin{equation}
H_{\textrm{eff}} = -\frac{K}{4 \pi} \left[({\bf b}_1 \cdot {\bf b}_2) \ln R - \frac{({\bf b}_1 \cdot {\bf R})({\bf b}_2 \cdot {\bf R})}{R^2}\right],
\end{equation}
where $\beta = 1/k_{\rm B}T$ and $K = [4 \mu (\mu + \lambda)]/(2 \mu + \lambda)$ is Young's modulus, which determines the interaction strength between dislocations and $R=|\mathbf{R}|$ is the distance  [see Fig. \ref{fig:fig3}(a)]. For a string of $N$ vacancies the distance can be written as ${\bf R} = (s,N \sqrt{3}/2)$ in the coordinate system where the $x$-axis is the direction of motion, and the Burgers vectors are given by ${\bf b}_1 = (-1,0)$, ${\bf b}_2 = (1,0)$. Using the dimensionless variable $x=2s/(\sqrt{3}N)$ one then obtains \cite{LECHNERDEFECTSTRINGS}
\begin{figure}[htb]
\centerline{\includegraphics[width=8.5cm]{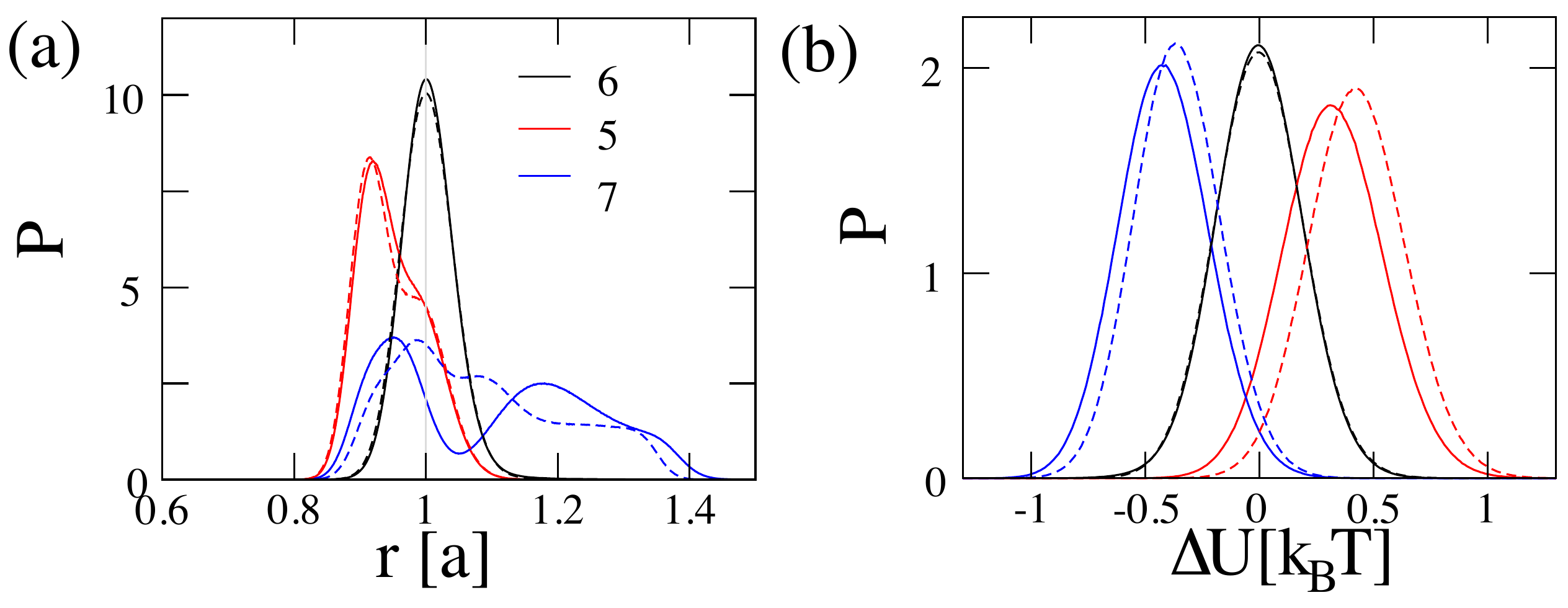}}
\caption{(a) Distribution of nearest neighbor distances for particles with $N_{nb}=6$ neighbors (black), $N_{nb}=5$ neighbors (red), and $N_{nb}=7$ neighbors (blue) at $r_d=65$ and $\kappa=0.05$ (solid) $\kappa=0.075$ (dashed). The distribution in the perfect crystal (black) is of Gaussian shape. Dislocations exhibit a more complicated and structured distribution of neighbor distances. Particles with $5$ neighbors show a peak at distances lower than a lattice distance. The distances for particles with $7$ neighbors are bimodal for low temperatures. This distribution can be understood from the microscopic picture of dislocations [see Fig. \ref{fig:fig1}(c)]. (b) Local potential energy of particles with $6$ (black), $5$ (red) and $7$ (blue) neighbors, respectively, relative to the average energy per particle in the system $\Delta U = U - U_0$. Note, that we consider the total energy, not just the nearest neighbor interactions.  The energy in the perfect crystal corresponds to a Gaussian distribution centered around zero. Particles with $7$ neighbors are locally in a lower energy environment as nearest neighbors are farther apart. Particles with $5$ neighbors have a larger energy, as neighboring particles have a smaller distance. The distributions separate well even for large temperatures $\kappa=0.075$ (dashed).}
\label{fig:fig4}
\end{figure}

\begin{equation}
\label{equ:endtoendscaled}
\log P(x) = -\beta \Delta F(x) = -\frac{K}{8 \pi} \left(\frac{1-x^2}{1+x^2} +  \ln \frac{1+x^2}{2} \right),
\end{equation}
 where the free energy has been shifted to vanish at $x=\pm1$. This function has the shape of a symmetric double well with minima at $x = \pm 1$. The minima are separated by a barrier at $x=0$ with height 

\begin{equation}
\label{equ:barrierheight}
h= \frac{K (1-\ln 2)}{8 \pi}.
\end{equation}

In the following we derive a critical value for $h$ below which the system enters the hexatic phase. Note, that the only fit parameter in the barrier height is the elastic constant $K$ . The critical value for $K$ below which the dislocations dissociate and the system enters the hexatic phase is $K<16 \pi$ \cite{FRENKEL}. This translates to a critical barrier height $h < h^\star= 0.63$. 

Fig. \ref{fig:fig3}(b) depicts the results for $r_d=65$ compared to $r_d=25$ and $\kappa = 0.054$. The barrier-height decreases with larger quantum fluctuations while keeping all other parameters fixed (from $h=0.79$ for $r_d=65$ to $h=0.54$ for $r_d=25$). This indicates a quantum hexatic phase ($h<h^\star$) induced purely by quantum fluctuations between the crystal and superfluid phase. The double-well interaction is measured from $P(x)$, the probability to find the string in a configuration with end-point separation $x$, where we used the vacancy positions in the string to follow the dislocation positions.

To detect dislocations in an experiment with polar molecules we propose the following protocol. Dislocations are defects with a larger local energy density as perfect crystalline regions. On a microscopic level this can be understood from the picture of dislocations as pairs of particles with $5$ and $7$ neighbors each [see Fig. \ref{fig:fig1}(c)]. Compared to particles in the perfect crystal, with $N_{nb}=6$ neighbors, these particles are in a different local environment from the interaction with their neighbors. Fig. \ref{fig:fig4}(a) depicts the distribution of nearest neighbors of particles in the perfect crystal compared to particles that form a dislocation defects. The difference in neighbor distances corresponds to a different potential energy of the particles depicted in Fig. \ref{fig:fig4}(b). 

It is this energy difference, which opens up a direct method to detect dislocations in an experiment with cold polar molecules:  the different rotational excited states exhibits a different dipole moment than the ground state, which may vanish for an optimal choice of parameters \cite{MICHELI}. Then, the application of a microwave field allows one to resolve the local energy of each polar molecules by measuring the number of rotationally excited polar molecules as a function of the frequency. As shown in Fig.\ref{fig:fig4}, the separation of these energies is large enough to resolve the the distributions even at high temperatures [see $\kappa=0.075$ Fig. \ref{fig:fig4}(b)].  Most promising is the combination of single site addressability \cite{GREINER} in a setup of polar molecules. Then, in a first step, interstitials can be created in a spatial resolved manner, and after equilibration, the position of the dislocation pairs at the end of the defect string are identified. Then, it is possible to generate in a first step a string of dislocations and afterwards directly measure the  separation probability shown in Fig.~\ref{fig:fig3}(b). These signatures allow for an alternative method to identify the hexatic phase beyond the analysis of the correlation functions.

\textit{Note added:} After completing this work we have become aware of  arXiv:1401.2237 (2014) G. M. Bruun and D. R. Nelson on quantum fluctuations in the hexatic phase from an analysis of Lindemann parameters. 

WL acknowledges support by the Austrian Science Fund (FWF): P 25454-N27. Work was supported by ERC Synergy Grant UQUAM, SFB FOQUS and EU-Projekt SIQS.

\bibliographystyle{prsty}

\section{Supplementary Material}

\subsection{Semi-classical Dynamics}

The dynamics of the system is evolved via a semi-classical path integral approach (Path Integral Langevin dynamics) introduced in Ref. \cite{PARINELLO}. This approach is similar to Path Integral Monte Carlo and takes advantage of the quantum to classical mapping of the partition function as Feynman path integrals \cite{FEYNMAN}. While in Monte Carlo, the paths are sampled via local update moves, the goal of Path Integral Langevin dynamics is to update the paths via molecular dynamics updates. The challenge of this approach is an efficient thermalization as low temperatures lead to large frequencies in the path integral an therefore small time-steps are required. This is solved by transforming into a normal mode picture and treat the center of mass mode differently as the high modes. In the following we repeat the derivation from \cite{PARINELLO}. The partition function of the system is   

\begin{equation}
\langle A \rangle = \frac{1}{Z}tr[e^{-\beta H} A],
\end{equation}
with 
\begin{equation}
Z = \frac{1}{(2 \pi \hbar)^f} \int d^f \mathbf{p} \int d^f \mathbf{q} e^{-\beta_n H_n(\mathbf{q},\mathbf{p})}.
\end{equation}
Here, $f=Nn$ and $\beta_n = \beta/n$. The full Hamiltonian consists of system Hamiltonian and interaction Hamiltonian

\begin{equation}
H_n(\mathbf{q},\mathbf{p}) = H_n^0(\mathbf{q},\mathbf{p}) + V_n(\mathbf{q}).
\end{equation}
The system part is 
\begin{equation}
H_n^0(\mathbf{q},\mathbf{p}) = \sum_{i=1}^{N}\sum_{j=1}^n \left(\frac{(p_i^{(j)})^2}{2 m_i} + \frac{1}{2}m_i \frac{1}{\beta_n^2 \hbar^2} [q_i^{(j)} - q_i^{j-1}]^2 \right)
\end{equation}

While sophisticated methods have been developed to sample this Hamiltonian with dynamics, an additional coupling to a thermal bath are more challenging as the frequencies in the path space separate from the frequency of the thermal bath. In Path Integral Langevin Dynamics, this is solved by propagating the Hamiltonian in the usual path integral space but the dissipative part of the thermal bath acts on the slowest mode of the normal modes of the path integral. This requires the normal mode represent of the Hamiltonian which is derived from the transformation
\begin{figure}[ht]
\includegraphics[width=8.0cm]{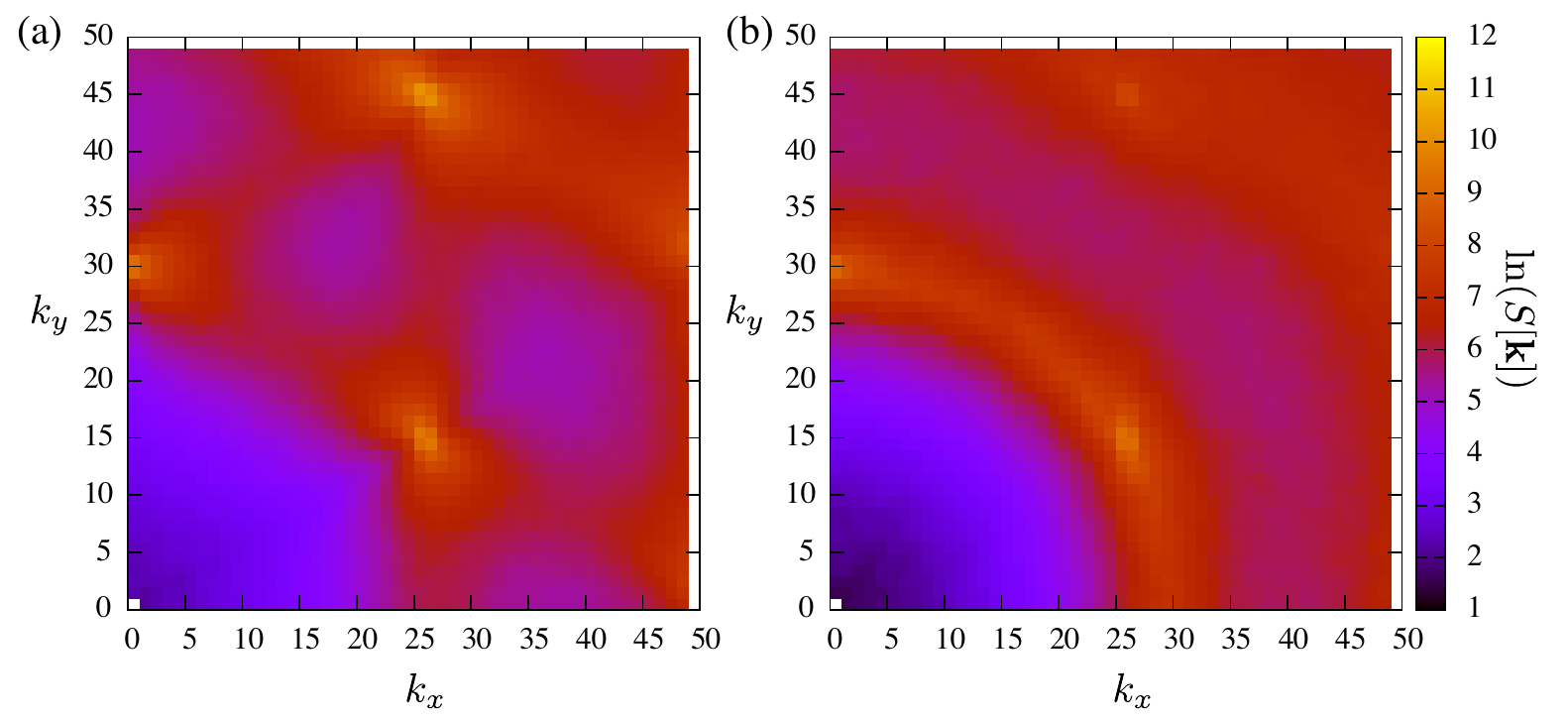}
\caption{The static structure factor $S(\bf{k})$ in the crystal phase $K=0.02$ and $r_d=20$ (a) and in the liquid phase with $K=0.14$ and $r_d=20$ (b). The crystal phase (a) features quasi-long-range order which leads to a pronounced maximum at the Bravis vector while the liquid phase ist structureless.  }
\label{fig:fig1}
\end{figure}

\begin{eqnarray}
\tilde{p}_k = \sum_{j=1}^n p_i^{(j)} C_{jk} \;  {\text and} \;\tilde{q}_k = \sum_{j=1}^n q_i^{(j)} C_{jk} 
\end{eqnarray}
where the matrix elements of $c_{jk}$ are

\begin{equation}
  C_{jk}=\begin{cases}
    \sqrt{1/n}, & \text{if $k=0$}\\
    \sqrt{2/n} \cos(2 \pi jk/n), & \text{if $1 \leq k \leq n/2-1$}\\
    \sqrt{1/n}(-1)^j, & \text{if $k = n/2$}\\
    \sqrt{2/n} \sin(2 \pi jk/n), & \text{if $n/2+1\leq k\leq n -1$}.
  \end{cases}
\end{equation}

In the normal mode representation the Hamiltonian reads as:

\begin{equation}
H_n^0(\mathbf{q},\mathbf{p}) = \sum_{i=1}^{N}\sum_{k=0}^{n-1} \left(\frac{(\tilde{p}_i^{(j)})^2}{2 m_i} + \frac{1}{2}m_i \frac{\sin(k \pi/n)}{\beta_n^2 \hbar^2} (\tilde{q}_i^{k})^2 \right)
\end{equation}
The Liouvillian is written as system part and interaction part. The coupling to the bath acts in the space of normal modes. 

\begin{equation}
L = L_0 + L_V.
\end{equation}
Here, $L_0 = -[H_n^0(\mathbf{q},\mathbf{p}),.]$ and $L_V = -[V_n^0(\mathbf{q},\mathbf{p}),.]$. 

\begin{equation}
e^{\Delta t L} = e^{(\Delta t/2)L_V}e^{\Delta t L_0}e^{(\Delta t/2)L_V}.
\end{equation}
While $L_0$ and $L_V$ are associated with momentum and position dynamics in the path integral the dissipative part is added in the equations of motion in the space of normal modes. This leads to the following equations of motion 

\begin{equation}
e^{\Delta t L} = e^{(\Delta t/2)L_\gamma}e^{(\Delta t/2)L_V}e^{\Delta t L_0}e^{(\Delta t/2)L_V}e^{(\Delta t/2)L_\gamma}
\end{equation}
While $L_0$ and $L_V$ are associated with momentum and position dynamics in the path integral, $L_\gamma$ is defined the space of normal mode variable. This requires four transformation between normal mode and path integral space per timestep. The step associated with $L_V$ is 

\begin{equation}
p_i^{(j)}(t + \Delta t) = p_i^{(j)}(t) - \Delta t \frac{\partial V}{\partial q_i^{(j)}}
\end{equation}
The evolution of the path integral with $L_0$ defined in normal mode space corresponds to 

\begin{eqnarray}
\tilde{p}_i^{(k)}(t + \Delta t) = \tilde{p}_i^{(j)}(t) \cos(\omega_k \Delta t) - \tilde{q}_i^{(j)}(t) m_i \omega_k \sin(\omega_k \Delta t). \\ \nonumber
\tilde{q}_i^{(k)}(t + \Delta t) = \tilde{p}_i^{(j)}(t)\frac{1}{m_i \omega_k} \sin(\omega_k \Delta t) - \tilde{q}_i^{(j)}(t) \cos(\omega_k \Delta t).
\end{eqnarray}
The last step, $L_\gamma$ describes thermalization in the normal mode space and leads to the step 

\begin{equation}
\tilde{p}_i^{(k)}(t + \Delta t) = \tilde{p}_i^{(k)}(t) e^{-\Delta t/2 \gamma^{(k)}} + \sqrt{\frac{m_i}{\beta_n} (1 - e^{-\Delta t \gamma^{(k)}})} \xi_i^{(k)}.
\end{equation}

\subsection{Spectroscopy}

The transition from the quasi-long range crystal phase to the hexatic phase requires the knowledge of the tails of the structure factor or, as described in the main text, the interaction between dislocations. The transition to the liquid phase, however, can be resolved with standard spectroscopy tools like Bragg scattering [see e.g. \cite{BRAGG}] from the static structure factor defined as

\begin{equation}
S(\mathbf{k}) = \frac{1}{N} \left \langle \sum_{jk} \mathrm{e}^{-i \mathbf{k} (\mathbf{R}_j - \mathbf{R}_k)} \right \rangle.
\end{equation}

The quasi-long-range nature of the correlation is clearly resolved from $S(\bf{k})$ with a sharp peak at the first Bravis vector [Fig. \ref{fig:fig1}(a)]. In the liquid phase no significant structures are left and $S(\bf{k})$ is structure-less [Fig. \ref{fig:fig1}(b)].

\bibliographystyle{prsty}

\end{document}